\documentclass[%
 reprint,
 amsmath,amssymb,
 aps,
]{revtex4-1}

\usepackage{graphicx}
\usepackage{dcolumn}
\usepackage{bm}

\begin{document}

\preprint{APS/123-QED}

\title{Two theorems for  the gradient expansion of relativistic hydrodynamics}

\author{Saulo Diles}
\email{smdiles@ufpa.br}
\affiliation{Campus Salin\'opolis,\\ Universidade Federal do Par\'a,\\
68721-000, Salin\'opolis, Par\'a, Brazil}

\begin{abstract}
This letter is dedicated to providing proof of two statements concerning the gradient expansion of relativistic hydrodynamics. The first statement is that \textit{the ordering of transverse derivatives is irrelevant in the gradient expansion of a non-conformal fluid}. The second statement is that \textit{the longitudinal projection of the Weyl covariant derivative can be eliminated in the gradient expansion of a conformal fluid}. This second statement does not apply to curvature tensors. In the conformal case, we know that the ordering of Weyl covariant derivatives is irrelevant in the gradient expansion.
\end{abstract}

\maketitle


 \section{Introduction}
 Hydrodynamics is one of the oldest fields of human knowledge and dates back to the third century B.C. \cite{johnson2016handbook}. Indeed, it is still a surprising and exciting field in theoretical (and experimental) physics.  The road to the formulation of a relativistic theory of fluid dynamics brought up new theoretical challenges. The causality problem in the relativistic version of Navier-Stokes equations and the definition of out-of-equilibrium thermodynamic variables \cite{eckart1940thermodynamics,hiscock1985generic,landau1959lifshitz,Bemfica:2017wps,Kovtun:2019hdm,Bemfica:2019knx,Bemfica:2020zjp} are important examples. To solve the former, higher-order derivatives have been introduced in equations of motion, establishing the M{\"u}ller-Israel-Stweart theory \cite{Muller:1967zza,Israel:1976tn,Israel:1979wp}.  A systematic formulation of relativistic hydrodynamics emerged as a gradient expansion \cite{Baier:2007ix,Bhattacharyya:2008jc,Romatschke:2009im}. The idea of a gradient expansion was firstly established as a second-order theory and then extended to  third order  \cite{Grozdanov:2015kqa,Diles:2019uft}. The definition of out-of-equilibrium quantities require a gauge fixing \cite{Kovtun:2019hdm,Bemfica:2019knx} that constrains the form of the conserved currents. 
 An in-depth analysis of the hydrodynamic frame choice led to a recent formulation of  hydrodynamics in a general frame \cite{Kovtun:2019hdm,Bemfica:2019knx,Bemfica:2020zjp,Hoult:2020eho,Noronha:2021syv}.
 Relativistic fluid dynamics is deeply connected with gravity through two different aspects: (i) the AdS/CFT duality and the fluid/gravity correspondence \cite{Policastro:2002se,kovtun2003holography,myers2009holographic,rangamani2009gravity,Hubeny:2010wp,Hubeny:2011hd,Campoleoni:2018ltl}, and (ii) the appearance of curvature tensors in second-order hydrodynamics. Within the holographic approach, and restricting to the linear response regime, the radius of convergence of the gradient expansion can be obtained in particular cases, such as the strongly coupled $\mathcal{N}=4$ SYM plasma \cite{Withers:2018srf, Grozdanov:2019kge, Grozdanov:2019uhi, Soltanpanahi:2021mys}.  
 In Ref.~\cite{Heller:2020uuy} it is shown that the convergence radius of the gradient expansion of conformal fluids is set by a momentum scale determined by the underlying microscopic theory. The results of Ref.~\cite{Heller:2020uuy} are independent of the existence of a holographic dual for the conformal fluid.
 
 Gradient expansion  is a methodology for establishing a theory of relativistic fluid dynamics. The first step is to identify the global conserved currents. In the general case, these are the energy-momentum tensor $T^{\mu\nu}$ and the matter current $J^\mu$. At order $n$, the conserved currents are expressed as a linear combination of all covariant terms that one can construct using the degrees of freedom and its derivatives, with at least $n$ derivatives in each term. A product of two first-order derivatives counts as a second-order gradient, and so on. Once this is done, one obtains the $n$-th-order currents, say  $T_n^{\mu\nu}$ and $J_n^\mu$. 
 
  Not all gradients   are independent. Dynamical equations and algebraic identities reveal linear dependencies in the set of gradients that one can include in a constitutive relation. For uncharged fluids, it was shown in \cite{Grozdanov:2015kqa} that only  transverse derivatives of $T,u^\alpha$ are relevant in the gradient expansion.  For conformal fluids, the gradient expansion can be performed by considering only Weyl-covariant derivatives of the degrees of freedom \cite{Bhattacharyya:2008mz,Loganayagam:2008is,Hubeny:2011hd,Diles:2017azi}. Covariance plays an important role. Only covariant derivatives can appear in the constitutive relations, ensuring the covariant nature of the conserved currents. One uses a connection to define a covariant derivative operator and apply this operator to the degrees of freedom. The connections play a special role as they can define curvature tensors from its ordinary derivatives ($\partial_\mu$). One cannot eliminate the longitudinal derivatives of curvature structures. 
 
 Here we extend the discussion on equivalences in the gradient expansion of ordinary and conformal fluids by proving two statements. This way, we establish two theorems for the gradient expansion of relativistic hydrodynamics. For ordinary fluids, we prove that the ordering of transverse derivatives is not relevant. For conformal fluids, we prove that the  Weyl covariant derivative of thermodynamic and mechanic degrees of freedom is transverse in all order of the gradient expansion.   
 This letter is organized as follows. In Section II, we discuss  the gradient expansion in a general situation. In section III, we discuss the case of an ordinary relativistic fluid and prove Theorem I. In Section IV, we discuss the relativistic conformal fluid and prove Theorem II. Section V is dedicated to discuss the consequences of the theorems proved here.

  \section{The  Gradient Expansion}

  The idea of gradient expansion is that the energy-momentum tensor and the matter current are given by a series expansion governed by the number of space-time derivatives acting on fundamental degrees of freedom.
  The ideal fluid establishes the zeroth-order theory:
  \begin{equation}
      T_0^{\mu\nu} = \epsilon u^\mu u^\nu + p\Delta^{\mu\nu},~~~J_0^\mu = n u^\mu.
  \end{equation}
In the above expression, we consider a mostly plus metric signature and use the transverse projector defined by $\Delta^{\mu\nu}=g^{\mu\nu}+u^\mu u^\nu$.
We then proceed by adding corrections to the ideal fluid constitutive relations. These corrections should be obtained by differentiating the fundamental degrees of freedom. As fundamental degrees of freedom, we consider the minimal set of independent functions featuring the fluid dynamics. For a fluid with one conserved charge, one can take $\{T, n, u^\alpha, g_{\mu\nu}\}$ as the fundamental degrees of freedom. The corrections we can include should preserve the covariance of the equations of motion. Thus, $T^{\mu\nu}$ corrections  are restricted to covariant symmetric rank-2 tensors, while the corrections to $J^\mu$ are restricted to covariant vectors. The large-scale symmetries of the system define covariance. Here we discuss two specific cases: the non-conformal relativistic fluid and the conformal relativistic fluid. Covariance also defines the equations of motion since the divergence-less of $T^{\mu\nu}$ and $J^\mu$ are defined by the appropriate covariant derivative.

The derivative order in each term is given by the total number of actions of the partial derivative, whether they act successively on the same function or separately on different functions.  At each order, we must look for all the different covariant structures that can be built. One general way to do that is proposed in Ref.~\cite{Grozdanov:2015kqa} for non-conformal fluids and extended in Ref.~\cite{Diles:2019uft} for conformal fluids.  Following such a procedure,
we find an extensive list containing all the
possible covariant terms at each order.  The $n$-th order correction in the gradient expansion is the linear combination of all order $n$ terms in the list. Each coefficient in the linear combination defines one transport coefficient. However, additional ingredients restrict this long list. The first ingredient is that defining non-equilibrium quantities requires gauge fixing, which strongly constrains the list of allowed corrections. We assume that a gauge choice is fixed from the beginning. The second ingredient concerns the algebraic equations relating gradients that are distinct \textit{a priori}. These equations include the zeroth-order equations of motion of the fluid and all the algebraic and geometric tensorial identities. Thus, one starts with the complete list of gradients allowed by the gauge fixing. For each additional equation relating two gradients, it is possible to replace one with the other and re-define the set of transport coefficients (see Ref.~\cite{Lahiri:2019lpk}).
Additional constraints on the transport coefficients arise when one imposes the second law of thermodynamics by constructing an entropy current. In this case, the constraints are not expressed as the vanishing of linear combinations of the transport coefficients, and they do not generate equivalences. For example, at first order, the positive divergence of the entropy current requires that $\zeta$ and $\eta$ are positive. More details concerning this discussion can be found in Refs.~\cite{Bhattacharyya:2012nq,Banerjee:2012iz,Jensen:2012jh,Bhattacharyya:2013lha,Bhattacharyya:2014bha} and references therein.
 
 For ordinary fluids, it has been shown in Ref.~\cite{Grozdanov:2015kqa} that the longitudinal projection of the gradients (curvature not included) is equivalent to a transverse derivative of a different degree of freedom. For this reason, in Ref.~\cite{Grozdanov:2015kqa} the list of corrections to the energy-momentum tensor contains only transverse gradients of $\ln s, u^\mu$. Here we extend this result by proving that the ordering of the transverse derivatives is irrelevant in the gradient expansion. For conformal fluids, a deep discussion on equivalences of the high order corrections is performed in Ref.~\cite{Loganayagam:2008is}, where the Weyl covariant derivative is established in the context of fluid dynamics. Here we prove that the longitudinal projections of the Weyl covariant derivative can be eliminated from the gradient expansion of a conformal fluid. This second theorem narrows the connections between non-conformal and conformal relativistic fluids.

 \section{The  non-conformal case: $\nabla_{\perp \mu}\nabla_{\perp \nu} \simeq \nabla_{\perp \nu}\nabla_{\perp \mu} $}

One convenient way to build the gradient expansion of an uncharged fluid is using only transverse covariant derivatives of temperature (or entropy) and velocity. Longitudinal gradients of such quantities are eliminated using the equations of motion. In the case of charged fluids, this extends to the concentration gradients.  In addition to these mechanic and thermodynamic gradients, we also consider corrections in the geometry using metric derivatives. For a torsion-free space-time, the covariant derivative of the metric vanishes, and covariant tensors involving metric derivatives come exclusively from the  Riemann tensor. 

Curvature tensors play a different role in gradient expansion. They arise from the commutation of covariant derivatives and are expressed in terms of connections. For a non-conformal system, the Riemann tensor and its contractions are the only curvature structures. In the gradient expansion of ordinary fluids, one cannot eliminate the longitudinal derivatives of the Riemann tensor.  The inclusion of the Riemann tensor and their covariant derivatives in the gradient expansion makes irrelevant the order of the transverse derivatives of the other degrees of freedom.

We assume that the gradient expansion of an ordinary fluid contains all the transverse gradients of $T, n, u^\mu$ and the covariant derivatives of the Riemann tensor $R_{\mu\nu~~\beta}^{~~~\alpha}$. In the case that one algebraic identity relates two or more of the referred tensors, say $\alpha T^I(\partial^n)+\beta F^I(\partial^n)=0$, one imposes that they are equivalent in the gradient expansion and write $T^I\simeq F^I$. Each equivalence allows removing one of the equivalent tensors from the gradient expansion. In this case, we say that this removed tensor is redundant and write $T^I\simeq 0$ if we choose $T^I$ to remove. To prove the irrelevance in the ordering of transverse derivatives, we first prove the Lemma below.

 {\bf Lemma 1:} For any equilibrium observable $\phi$,  if the longitudinal projection of the first-order derivative is redundant, $u^\alpha\nabla_\alpha\phi\simeq 0$, then the second-order longitudinal derivative is also redundant: $u^\mu\nabla_\mu(u^\alpha \nabla_\alpha\phi)\simeq 0$.  
 
 {\bf Proof:} The equations of motion of the ideal charged fluid are \cite{Muronga:2003ta,Lahiri:2019lpk}
 \begin{align}
     u^\alpha\nabla_\alpha \ln s &= - \nabla_{\perp \alpha}u^\alpha,  \label{entropy}\\
     u^\alpha\nabla_\alpha u^\mu & = - \frac{1}{sT+\mu n} \nabla_\perp^{~\mu}p, \label{vel}  \\
     u^\alpha\nabla_\alpha \ln n &=  - \nabla_{\perp \alpha}u^\alpha.\label{charge}
 \end{align}
The transverse gradient is defined as $\nabla_{\perp \alpha} = \Delta_\alpha^{~\beta}\nabla_\beta$. For simplicity, write the right-hand side of eq.~(\ref{vel}) in terms of pressure gradients. For scalar degrees of freedom we have that $D\phi_{\text{scalar}}=-\nabla_{\perp \alpha}u^\alpha\simeq 0$, where $D\equiv u^\alpha\nabla_\alpha$ defines the longitudinal derivative. For  the second-order longitudinal derivative, one has
 \begin{align}
    D(D\phi_{\text{scalar}})&=- D(\nabla_{\perp \mu}u^\mu) \nonumber \\
    &= - u^\alpha\nabla_\alpha( \nabla_{ \mu}u^\mu) \simeq u^\alpha\nabla_\mu\nabla_\alpha u^\mu \nonumber \\
    &= - (\nabla_\mu u_\alpha)\nabla^\alpha u^\mu + \nabla_\mu (Du^\mu).
 \end{align}
The first term is transverse and the second term requires more attention. We have from eq.~(\ref{vel}) that $\nabla_\mu(Du^\mu) \simeq \nabla_\mu(\nabla_\perp^{~\mu}p)$. However, the divergence of a transverse gradient is also transverse:
\begin{align}
    \nabla_\mu(\nabla_\perp^{~\mu}p) &= \nabla_\mu\nabla^\mu p + (\nabla_\mu u^\mu )Dp \nonumber \\
   &+ u_\mu(\nabla^\mu u^\alpha)\nabla_\alpha p + u^\mu u^\alpha \nabla_\mu \nabla_\alpha p.
\end{align}
The second term in the above equation is transverse and the third therm is redundant: it is equivalent to $\nabla_{\perp}^{\mu}p\nabla_{\perp \mu}p$. The first and the last terms combine into a transverse Laplacian and we have
\begin{equation}
\nabla_\mu(\nabla_\perp^{~\mu}p) 
\simeq \nabla_{\perp \mu}\nabla_\perp^{~\mu}p\simeq 0.
\end{equation}
 Consequently, 
 \begin{equation}
     D(D\phi_{\text{scalar}}) \simeq 0.
 \end{equation}
 For the velocity field, we have that
 \begin{align}
     D(Du^\mu) &\simeq u^\alpha\nabla_\alpha(\nabla_\perp^{~\mu}p)\nonumber \\
         &\simeq u^\alpha\nabla_\alpha(\nabla^{~\mu}p) + u^\mu u^\alpha D(\nabla_\alpha p) \nonumber \\
     &= \nabla^\mu(Dp) - (\nabla^\mu u^\alpha)\nabla_\alpha p \nonumber \\
         &+ u^\mu D(Dp) -u^\mu(Du^\alpha)\nabla_\alpha p \nonumber \\
         &\simeq \nabla^\mu (Dp)+ u^\mu D(Dp)=\nabla_\perp^\mu(Dp)\nonumber \\
         &\simeq \nabla_\perp^\mu (\nabla_{\perp \alpha}u^\alpha)\simeq 0.
 \end{align}
 The equivalence in the first line comes only from eliminating products of first-order gradients, whereas in the last line eqs. (\ref{entropy}) and (\ref{charge}) were also used.
 
 {\bf Theorem 1:} The ordering of transverse derivatives is irrelevant in the gradient expansion of a non-conformal relativistic fluid. For any equilibrium observable $\phi$ we have that $\nabla_{\perp \mu}\nabla_{\perp \nu}\phi \simeq \nabla_{\perp \nu}\nabla_{\perp \mu}\phi$.

 {\bf Proof:}  To show that the ordering of transverse derivatives is irrelevant, we compute the commutator of two transverse derivatives acting on an arbitrary degree of freedom. Let $\phi$ be $T, n$ or $u^\alpha$, so that $\phi$ satisfies $u^\mu\nabla_\mu\phi\simeq 0$.
 Then, one can compute the commutator $[\nabla_{\perp \mu},\nabla_{\perp \nu}]\phi$:
 
 \begin{align}
     [\nabla_{\perp \mu},\nabla_{\perp \nu}]\phi &=  \Delta_\mu^{~\alpha}\nabla_\alpha (\Delta_\nu^{~\beta}\nabla_\beta \phi) - 
      \Delta_\nu^{~\alpha}\nabla_\alpha (\Delta_\mu^{~\beta}\nabla_\beta \phi) \nonumber \\
       &=  [\nabla_\mu,\nabla_\nu]\phi + u_\mu u^\alpha\nabla_\alpha(\nabla_\nu\phi+u_\nu u^\beta\nabla_\beta\phi) \nonumber \\
  &- u_\nu u^\alpha\nabla_\alpha(\nabla_\mu\phi+u_\mu u^\beta\nabla_\beta\phi)
 \end{align}
 
The first term in the above equation is a commutator of covariant derivatives and is given by a linear combination of contractions with the Riemann tensor. Since all the Riemann tensors and their contractions are already included in the gradient expansion, one has that $[\nabla_\mu,\nabla_\nu]\phi\simeq 0$. This leads to
\begin{align}
    [\nabla_{\perp \mu},\nabla_{\perp \nu}]\phi &\simeq  u_\mu u^\alpha\nabla_\alpha(\nabla_\nu\phi+u_\nu u^\beta\nabla_\beta\phi) \nonumber \\
  &- u_\nu u^\alpha\nabla_\alpha(\nabla_\mu\phi+u_\mu u^\beta\nabla_\beta\phi) \nonumber \\
  &= u_\mu u^\alpha\nabla_\alpha\nabla_\nu\phi + u_\mu u_\nu u^\alpha u^\beta \nabla_\alpha\nabla_\beta\phi \nonumber \\
  &+u_\mu u^\alpha u^\beta (\nabla_\alpha u_\nu)\nabla_\beta\phi + u_\mu u_\nu u^\alpha(\nabla_\alpha u^\beta)\nabla_\beta \phi \nonumber \\
  &-u_\nu u^\beta\nabla_\beta\nabla_\mu\phi - u_\nu u_\mu u^\beta u^\alpha \nabla_\beta\nabla_\alpha \phi \nonumber \\
  &-  u_\nu u_\mu u^\beta (\nabla_\beta u^\alpha) \nabla_\alpha\phi - u_\nu u^\beta u^\alpha (\nabla_\beta u_\mu)\nabla_\alpha\phi \nonumber \\ 
  &= u_\mu u^\alpha\nabla_\alpha\nabla_\nu\phi +u_\mu u^\alpha u^\beta (\nabla_\alpha u_\nu)\nabla_\beta\phi  \nonumber \\
  & + u_\mu u_\nu u^\alpha(\nabla_\alpha u^\beta)\nabla_\beta \phi \nonumber \\
  &-u_\nu u^\beta\nabla_\beta\nabla_\mu\phi -     u_\nu u_\mu u^\beta (\nabla_\beta u^\alpha) \nabla_\alpha\phi \nonumber \\
  & - u_\nu u^\beta u^\alpha (\nabla_\beta u_\mu)\nabla_\alpha\phi. \label{commute}
\end{align}
 In the last line, it was used the fact that $ u^\alpha u^\beta [\nabla_\alpha,\nabla_\beta]\phi=0$. We proceed by analyzing the remaining terms.
 The second and the sixth terms are of the same kind, so we discuss explicitly only the second one. We have that $u_\mu u^\alpha u^\beta (\nabla_\alpha u_\nu)\nabla_\beta\phi = u_\mu ( u^\alpha  \nabla_\alpha u_\nu)   u^\beta  \nabla_\beta\phi\simeq 0$, since the longitudinal derivatives can be eliminated. The third and the fifth terms are also of the same kind and we discuss explicitly the third one. We notice first that  $(\nabla_\alpha u^\beta)\nabla_\beta \phi = (\nabla_\alpha u^\beta)\nabla_{\perp \beta} \phi$ as a consequence of $u_\beta u^\beta=-1$. Then we have that $u_\mu u_\nu u^\alpha(\nabla_\alpha u^\beta)\nabla_\beta \phi =
 u_\mu u_\nu (u^\alpha\nabla_\alpha u^\beta) \nabla_{\perp \beta} \phi$. Nevertheless the longitudinal derivatives of the velocity can be replaced by transverse derivatives. Thus, this term is equivalent to one involving only transverse gradients  already in the list. We have that $u_\mu u_\nu u^\alpha(\nabla_\alpha u^\beta)\nabla_\beta \phi\simeq 0$.
 
 We remain with the first and fourth terms in eq.~(\ref{commute}), obtaining
 \begin{equation}
     [\nabla_{\perp \mu},\nabla_{\perp \nu}]\phi \simeq u_\mu u^\alpha\nabla_\alpha\nabla_\nu\phi  -u_\nu u^\beta\nabla_\beta\nabla_\mu\phi.
 \end{equation}
 The two terms are of the same kind, differing by index commutation. Thus we analyze the first one explicitly. At this stage, it is convenient to change the ordering of covariant derivatives, yielding
  \begin{equation}
     [\nabla_{\perp \mu},\nabla_{\perp \nu}]\phi \simeq u_\mu u^\alpha \nabla_\nu \nabla_\alpha\phi  -u_\nu u^\beta\nabla_\mu \nabla_\beta\phi.
 \end{equation}
 Finaly, we use that as $u^\alpha \nabla_\alpha \phi \simeq 0$ then $\nabla_\alpha \phi \simeq \nabla_{\perp \alpha}\phi$. From Lemma 1, we have that $ \nabla_\nu (\nabla_\alpha\phi) \simeq \nabla_\nu( \nabla_{\perp \alpha}\phi)$. We can now differentiate the equality $u^\alpha\nabla_{\perp \alpha}\phi=0$, obtaining  
 \begin{equation}
    u^\alpha \nabla_\nu (\nabla_{\perp \alpha}\phi) \simeq -(\nabla_{\perp\nu } u^\alpha )\nabla_{\perp \alpha} \phi \simeq 0.  
 \end{equation}
Using the above equation we can conclude that $u_\mu u^\alpha \nabla_\nu \nabla_\alpha\phi \simeq 0$. Hence
\begin{equation}
     [\nabla_{\perp \mu},\nabla_{\perp \nu}]\phi \simeq 0. 
\end{equation} 
 
 \section{The  Weyl covariant gradient expansion}
 
 A Weyl transformation \cite{Weyl:1918ib} is a local scaling of the line element $ds\to e^{\Phi(x)}ds$, equivalent to a redefinition of the metric tensor $g_{\mu\nu}\to e^{2\Phi}g_{\mu\nu}$. A tensor $\mathcal{T}$ is said to be of weight $\omega$ under Weyl scaling if it transforms as $\mathcal{T}\to e^{-\omega\Phi}\mathcal{T}$.  Scaling invariance, or Weyl invariance, is expected to manifest in the hydrodynamic limit of a conformal field theory. As a consequence, the constitutive relations of such a fluid should be compatible with this (local) scaling invariance.  The compatible energy-momentum tensor, $T^{\mu\nu}$, transforms as a rank-2 tensor density of weight $d+2$ under Weyl scaling, where $d$ is the number of space-time dimensions. The compatible vector current, $J^\mu$,   transforms as a contravariant vector density of weight $d$. 
 
  The fundamental degrees of freedom in hydrodynamics are tensor densities under Weyl scaling. Thus, we need to ensure that their derivatives will also be tensor densities. To do so, a derivative operator that is covariant under Weyl transformations is needed. This operator is the Weyl covariant derivative \cite{Loganayagam:2008is}.    The   Weyl covariant derivative of a field $\psi$ with weight $\omega$, denoted here by $\mathcal{D}_\alpha\psi$,  is obtained by replacing everywhere $\partial_\alpha \psi\to (\partial_\alpha+\omega\mathcal{A}_\alpha)\psi$, where  $\mathcal{A}_\alpha$ is the Weyl connection  \cite{Diles:2017azi}.  Let $\Phi$ be a scalar with weight $\omega_\Phi$ and $\gamma^\alpha$ a vector with weight $\omega_\gamma$.  The associated Weyl covariant derivatives are given respectively by
 \begin{align}
    \mathcal{D}_\alpha \Phi &= (\partial_\alpha +\omega_\Phi\mathcal{A}_\alpha)\Phi, \nonumber \\
     \mathcal{D}_\beta \gamma^\alpha  &=(\partial_\beta +\omega_\gamma\mathcal{A}_\beta)\gamma^\alpha +\bar{\Gamma}^{~\alpha}_{\beta\mu} \gamma^\mu,
 \end{align}
 where 
 \begin{align}
   \bar{\Gamma}^{~\alpha}_{\beta\mu} &=    \frac{1}{2}g^{\alpha\nu} [(\partial_\beta - 2\mathcal{A}_\beta)g_{\nu\mu} \nonumber \\ &+(\partial_\mu - 2\mathcal{A}_\mu)g_{\nu\beta} - (\partial_\nu - 2\mathcal{A}_\nu)g_{\beta\mu}]
 \end{align}
 is obtained by performing $\partial g_{\alpha\beta}\to (\partial - 2\mathcal{A})g_{\alpha\beta}$ in the Christoffel symbol. 
 
 There is a unique way to define the Weyl connection in hydrodynamics, which is requiring it to be transverse and trace-less when acting on the velocity field. These   conditions are expressed as
 \begin{equation}
     \mathcal{D}_\alpha u^\alpha=0,~u^\alpha\mathcal{D}_\alpha u^\beta=0.  \label{derive}
 \end{equation}
 The former leads to $u^\alpha\mathcal{A}_\alpha=-\frac{\nabla_\alpha u^\alpha}{(d-1)}$, while the latter gives  $\mathcal{A}^\alpha = u^\beta\nabla_\beta u^\alpha + u^\beta\mathcal{A}_\beta u^\alpha$. We then obtain the unique Weyl connection satisfying both requirements:
\begin{equation}
    \mathcal{A}_\alpha = u^\beta\nabla_\beta u_\alpha - \frac{(\nabla_\beta u^\beta)}{d-1}u_\alpha.
\end{equation} 
The above definition of the Weyl connection ensures that eq.~(\ref{derive}) holds even in the off-shell case.  For a conformal fluid, transversality of the Weyl covariant derivative of the velocity field is not an equivalence but an equality. Due to the trace-less condition, we have for the scalars degrees of freedom that
\begin{equation}
    u^\alpha\mathcal{D}_\alpha T = 0,~~    u^\alpha\mathcal{D}_\alpha \ln n =0.
\end{equation}

In the conformal case, we also find that the longitudinal derivative redundancy does not apply to curvature tensors.  The condition that the covariant derivative of the metric vanishes is not affected by the Weyl connection. We have 
\begin{equation}
    \mathcal{D}_\alpha g_{\mu\nu} =0.
\end{equation}
This does not mean that metric derivatives are excluded from the gradient expansion. Metric derivatives and Weyl connection derivatives are encoded in the curvature tensors $\mathcal{R}_{\alpha\beta~\sigma}^{~~~\mu}$ and $\mathcal{F}_{\alpha\beta}$. The former is the conformal Riemann tensor, while the latter is the field strength for the Weyl connection.  They appear in the commutator of Weyl covariant derivatives, which acts on a vector field as
\begin{equation}
    [  \mathcal{D}_\alpha, \mathcal{D}_\beta]\xi^\mu = -\omega_\xi \mathcal{F}_{\alpha\beta}\xi^\mu - \mathcal{R}_{\alpha\beta~\sigma}^{~~~\mu}\xi^\sigma.\label{Dcommut}
\end{equation}

In the following, we show that transversality of the Weyl covariant derivative extends to higher-order terms in the gradient expansion.

{\bf Theorem 2:} At any order of gradient expansion  of a conformal fluid, the longitudinal projection of the Weyl covariant derivative of entropy, concentration, and velocity are redundant. 

{\bf Proof:} Let $\phi_i$ be either temperature, concentration, or a velocity component. It holds that
\begin{equation}
    u^\alpha\mathcal{D}_\alpha\phi_i=0\to \mathcal{D}_\alpha\phi_i = \mathcal{D}_{\perp \alpha}\phi_a,
\end{equation}
 where  $\mathcal{D}_{\perp \alpha}\equiv\Delta_\alpha^{~\beta}\mathcal{D}_\alpha$. We now consider the second order derivative:
\begin{equation}
    \mathcal{D}_\beta \mathcal{D}_\alpha \phi_i =   \mathcal{D}_\beta (\mathcal{D}_{\perp \alpha} \phi_i). \label{conftrans}
\end{equation}
We first note that by differentiating $u^\alpha\mathcal{D}_{\perp \alpha} \phi_i=0$,  we get
\begin{equation}
    u^\alpha \mathcal{D}_\beta \mathcal{D}_\alpha \phi_i =-( \mathcal{D}_\beta u^\alpha)\mathcal{D}_\alpha \phi_i 
    \simeq 0. \label{2equiv}
\end{equation}
 From eq.~(\ref{Dcommut}) we obtain that $    u^\alpha \mathcal{D}_\beta \mathcal{D}_\alpha \phi_i \simeq  u^\alpha  \mathcal{D}_\alpha \mathcal{D}_\beta \phi_i = u^\alpha  \mathcal{D}_\alpha \mathcal{D}_{\perp \beta}  \phi_i$. 
 We now use eq.~(\ref{2equiv}) to obtain
 \begin{equation}
  u^\alpha  \mathcal{D}_\alpha \mathcal{D}_\beta \phi_i  \simeq
  -(\mathcal{D}_\beta u^\alpha)\mathcal{D}_\alpha \phi_i 
    \simeq 0.
 \end{equation}
 In the gradient expansion of a conformal fluid, if the first derivative is transverse, the second will also be transverse. The arguments above are extended to higher-order derivatives by taking $\phi$ as a lower-order transverse gradient. 
 Consequently, in any order of gradient expansion, only the transverse derivatives of the degrees of freedom  will be present and give rise to new transport coefficients. In other words, the Weyl covariant derivative of any hydrodynamic degree of freedom is always transverse, despite the order of gradient expansion.

 \section{Discussion}

Modern understanding of relativistic hydrodynamics employs a gradient expansion. In this context, a systematic way of generating  arbitrarily high order corrections of the constitutive relations is needed. The present work paves one more step in establishing the rules to fix constitutive relations and the number of transport coefficients of a viscous fluid. Our present results hold in the general, non-linear case.
 
 The statements proved here lead to consequences in the building of the gradient expansion of relativistic fluids. It is important to mention that in Ref.~\cite{Diles:2019uft},  the statement of  Theorem 1 has been conjectured to hold in the general case and proved to hold only in some particular cases. The consequences of Theorem 1 have already been explored in that reference: it leads to a reduction of 10 in the number of transport coefficients for third-order hydrodynamics of uncharged fluids. We also remark that in the inspiring work presented in Ref.~\cite{Grozdanov:2015kqa}, it is proved that only the transverse derivatives are relevant in the gradient expansion. However, the authors of that reference did not raise the possibility that commuting transverse derivatives would give rise only to redundant structures. The effects of the irrelevance in the ordering of transverse derivatives in higher-order constitutive relations is to decrease the number of transport coefficients.  
 
 The consequences of Theorem 2 have not been explored in higher-order hydrodynamics. This theorem states that the Weyl covariant derivative acts on mechanic and thermodynamic degrees of freedom of the fluid as a transverse operator in the context of gradient expansion. Consequently, it suggests that  therms like $u^\alpha\mathcal{D}_\alpha\sigma^{\mu\nu}$ or $u^\alpha\mathcal{D}_\alpha \Omega^{\mu\nu}$ can, in principle, be eliminated from the gradient expansion and its transport coefficients absorbed into the remaining ones. Here the tensors $\sigma^{\mu\nu}$ and $\Omega^{\mu\nu}$ correpond to the symmetric and anti-symmetric parts of $\mathcal{D}^\mu u^\nu$, respectively.  However, it is not sufficient to state the redundance of these structures. What happens is that  the equality
  \begin{align}
      u^\alpha\mathcal{D}_\alpha\sigma^{\langle\mu\nu\rangle} =& -\sigma^{\alpha\langle\mu}\sigma_{~~\alpha}^{\nu\rangle} - \Omega^{\alpha\langle\mu}\Omega_{~~\alpha}^{\nu\rangle}\nonumber \\
      & + u^\alpha u^\beta \mathcal{R}_{\alpha~~~~\beta}^{~\langle\mu\nu\rangle}, \label{longsig}
  \end{align}
  is satisfied. Notice that in the expression above $\mathcal{R}_{\alpha~~\beta}^{~\mu\nu}$ corresponds to the curvature tensor for the Weyl-covariant derivative defined in eq.~(\ref{Dcommut}) and $T^{\langle\mu\nu\rangle}= \frac{1}{2}\Delta^{\mu\alpha}\Delta^{\nu\beta}(T_{\alpha \beta}+ T_{\beta\alpha}) - \frac{2}{d-1}\Delta^{\mu\nu}\Delta_{\alpha\beta}T^{\alpha\beta}$ is the transverse, symmetric, and traceless projection. Thus, the longitudinal derivative $u^\alpha\mathcal{D}_\alpha\sigma^{\langle\mu\nu\rangle}$ will be redundant only if eq.~(\ref{longsig}) is linearly independent of all other algebraic relations  used to reduce the list of second-order gradients. On the other hand, $u^\alpha\mathcal{D}_\alpha \Omega^{\langle\mu\nu\rangle}$ vanishes identically due to the anti-symmetry of $\Omega^{\mu\nu}$.

It is important to remark that an old result of the fluid/gravity correspondence \cite{Bhattacharyya:2008jc,Haack:2008cp,Bhattacharyya:2008mz} states that the coefficient $\lambda_3$ associated with the tensor $\Omega^{\alpha\langle\mu}\Omega_{~~\alpha}^{\nu\rangle}$ vanishes for  $\mathcal{N}=4$ SYM conformal fluids in the limit of large $N_c$ and large t'Hooft coupling $\lambda$. However, finite $\lambda$ corrections result in a non-vanishing $\lambda_3$, as shown in Ref.  \cite{Saremi:2011nh, Grozdanov:2014kva}. The results of Ref.~\cite{Saremi:2011nh, Grozdanov:2014kva} hold for a specific case and even  suggesting that there are, in fact, five independent second-order conformal gradients, that is not enough to conclude that this is valid in the general case.  A detailed  discussion on the possibility of reducing the list of second-order conformal gradients is needed and will be carried over in  future work.

\begin{acknowledgments}
The author thanks Alex S. Miranda, the first who noticed that changing the ordering of transverse derivatives does not generate new structures. This observation led to Theorem 1 as a conjecture and motivated the search for its proof. The author thanks the Campus Salinopolis of the Universidade Federal do Par\'a for the release of work hours for research.

\end{acknowledgments}

\appendix

 \bibliographystyle{ieeetr}
\bibliography{RevisedManuscript.bib}  

\end{document}